# Modeling Engagement Signals in Technology-Enhanced Collaborative Learning: Toward AI-Ready Feedback

Joan (Ruiqiong) Zhong[1, *]

[1]Menlo College, California, USA

[*]Corresponding author: rqzhong1@gmail.com

## Abstract

Modeling engagement in collaborative learning remains a persistent challenge, particularly in technology-enhanced environments where surface indicators—such as participation counts or speaking frequency—are easily misinterpreted. This study proposes a lightweight, interpretable, and classroom-embedded framework that operationalizes shared understanding (Q2) and consensus building (Q4) as observable proxies of collaborative engagement processes. Grounded in learning sciences and inspired by mechanisms identified in neuroscience research, these proxies were consolidated into a Composite Signal Index (CSI). Combined with sustained motivation (Q6), CSI supports a quadrant diagnostic model that differentiates engagement states with implications for teacher–AI collaborative feedback.

An exploratory validation was conducted in an adult ESL classroom using a structured three-phase collaborative speaking task (rotating reading → retelling → consensus). Results indicated a positive association between CSI and sustained motivation, while qualitative reflections highlighted the regulatory value of peer feedback (Q3) as a candidate for future NLP-based modeling. The study further demonstrates an AI-ready prototype, illustrating how structured signals can be mapped onto transparent diagnostic logic and differentiated instructional prompts.

By translating mechanism-informed indicators into reproducible classroom signals, this study provides a scalable and equitable approach to engagement modeling. The model enhances equitable recognition of diverse learners by distinguishing mechanism-aligned structural engagement signals from surface participation behaviors. The framework emphasizes that silence does not equal disengagement and that frequent talk does not guarantee cognitive depth, helping teachers recognize marginalized learners who often remain invisible in surface-level analytics. The findings outline a "Bridging Stage" from exploratory feasibility toward future deployment in technology-enhanced learning systems.

# 1. Introduction

## 1.1 Background and Rationale

Learner engagement is widely recognized as a decisive factor in both personalized and collaborative learning, directly shaping participation, motivation, and achievement. Classic frameworks conceptualize engagement as comprising behavioral, emotional, and cognitive dimensions (Fredricks et al., 2004). Instruments such as the Student Engagement Instrument (SEI; Appleton et al., 2006) provide structured means to assess psychological and cognitive engagement at the individual level. Yet, in collaborative learning, engagement emerges not as an isolated state but as a socially distributed and co-regulated process—one that unfolds dynamically through mutual understanding, responsive feedback, and consensus building.

**Educational need.** Existing classroom indicators—such as participation counts or speaking duration—are surface-level and prone to misclassification: frequent talk does not guarantee deep thinking, and silence does not imply disengagement (Thurn et al., 2023). Even behavior-only prediction models risk confusing deep concentration with low engagement while over-valuing high-frequency talk (Apicella et al., 2022; Karimah & Hasegawa, 2022). Multimodal techniques, including EEG, eye-tracking, and facial analysis, can achieve high accuracy in controlled settings but remain costly, intrusive, and impractical for everyday classrooms. This persistent gap underscores the need for lightweight, interpretable, and scalable proxies that can be embedded into routine instruction. Meanwhile, surveys, observations, and behavioral logs still fall short in capturing collaborative dynamics. Critical phenomena such as shared understanding (Q2), feedback responsiveness (Q3), and consensus formation (Q4) remain under-represented in current models, leaving silent or marginalized learners easily overlooked (Holstein & Doroudi, 2021).

**Neuroscience-informed potential.**

Research across educational neuroscience has shown that episodes of shared understanding and consensus often coincide with speaker–listener neural coupling or group-level synchrony, offering mechanism-level insight into why these events matter for collaborative learning (Stephens et al., 2010; Pan et al., 2020).

**Cross-level methodological contribution.**

Building on this insight, the present study treats Q2 (shared understanding) and Q4 (consensus) as observable behavioral counterparts of these mechanism-level events and integrates them into a Composite Signal Index (CSI). This cross-level pathway—from neuroscientific insight to behavioral proxies, to CSI, and finally to CSI × Q6 interpretation—constitutes the study's methodological contribution. It enables collaborative learning to be examined through a lightweight and interpretable diagnostic lens inspired by hyperscanning research, without relying on neural measurement.

**Focus of the present study.**

Accordingly, the study focuses on learners' observable acts of meaning-making and negotiation during collaboration rather than teacher–student interaction. Moments of shared understanding (Q2) and consensus formation (Q4) are treated as mechanism-informed behavioral events that can be translated into classroom-deployable, AI-interpretable signals for modeling individual engagement states within collaborative tasks.

This study therefore aims to provide a mechanism-informed, classroom-feasible, and equity-aware pathway for identifying individual engagement within collaborative tasks—an aspect of TEL that remains substantially underdeveloped.

## 1.2 Research Problem and Challenges

A central challenge in collaborative learning is determining which learners are genuinely engaging in joint meaning-making and cognitive effort, rather than merely participating on the surface. Three major issues hinder current engagement modeling:

**1) Individual-level bias.**

Most AI-driven engagement models rely on individual behavioral traces—such as click rates, dwell time, or video operations—while overlooking the interactional, regulatory, and motivational dynamics that shape engagement in collaborative settings. Such individual proxies are often context-dependent and limited in generalizability (Kuhlman et al., 2024). Many tools fail to capture cognitive regulation (Panadero et al., 2023), and collaborative phenomena such as consensus formation and shared goals are frequently neglected (Kovari, 2025; Xu et al., 2024). As a result, existing methods struggle to represent the interaction-informed engagement states that emerge during collaboration.

**2) Neural–behavioral disconnect.**

Although brain-to-brain synchrony (BBS) studies highlight the importance of shared understanding and consensus in collaborative learning (Pan et al., 2023), these findings rely on EEG/fNIRS hyperscanning in controlled laboratory conditions. They offer theoretical insight but provide few scalable, classroom-ready pathways for detecting such mechanisms in everyday instruction.

**3) Scalability and equity limitations.**

Many classroom analytics systems still emphasize surface-level activity counts and lack sensitivity to the subtle interaction processes experienced by different learners. Social engagement indicators may even show negative effects depending on group dynamics (Xing et al., 2022). Silent or marginalized learners are easily overlooked, raising concerns about equity (Holstein & Doroudi, 2021).

In response to these challenges, this study adopts a design–test–reflection logic and conducts an exploratory validation in authentic ESL classrooms. Drawing on mechanism-informed signals—Q2 (shared understanding), Q4 (consensus), and Q6 (sustained motivation)—it proposes the CSI × Q6 framework as a lightweight, interpretable pathway for modeling individual engagement within collaborative tasks and supporting AI-ready feedback.

## 1.3 Research Entry Point and Methodological Pathway

To bridge these gaps, this study introduces a cross-level translational pathway: from neuroscience insights, to classroom-observable signals, and finally to AI-ready engagement modeling. At the behavioral level, Q2 (shared understanding) and Q4 (consensus) are consolidated into a Composite Signal Index (CSI). This index is then combined with Q6 (sustained motivation) to form a CSI × Q6 quadrant framework that distinguishes group states and informs feedback design. This study follows a translational pathway: from synchrony-informed mechanisms to classroom-observable behavioral proxies, and from proxy-level interpretation to AI-ready feedback.

A small-scale ESL classroom study was conducted with three-person collaborative speaking tasks (rotating reading, retelling, feedback, consensus). This structured design naturally elicited understanding and consensus behaviors, enabling empirical validation of the CSI and its linkage with motivation. This pathway thus operationalizes neuroscientific mechanisms into structured and reproducible signals, and further positions these signals as foundational inputs for AI-enabled prototype design to support real-time feedback.

This study is not a neuroscientific experiment; rather, it draws on neural findings as inspiration to design a deployable modeling pathway for educational AI.

## 1.4 Research Contributions

This study contributes to engagement research in three interconnected ways. It advances theory by translating neuroscience-inspired findings on synchrony into classroom-observable signals of shared understanding and consensus. Empirically, it demonstrates through a proof-of-concept ESL classroom study that these structured signals can be integrated into a Composite Signal Index (CSI) and meaningfully related to sustained motivation. Practically, it outlines an AI-ready framework that is lightweight, interpretable, and equity-aware, offering a pathway for teacher–AI complementarity in collaborative learning contexts.

## 1.5 Research Questions

Accordingly, this study investigates:

- RQ1. Can key mechanisms revealed in neuroscience (understanding and consensus) be translated into classroom-observable signals?

- RQ2. Can the Composite Signal Index (CSI) relate to sustained motivation and distinguish different group engagement states?

- RQ3. Can these structured signals serve as AI-ready inputs, providing a feasible pathway for real-time feedback in collaborative learning?

## 2. A Neuro-Inspired Pathway for Observable Signals in Collaborative Engagement

This section reviews key challenges in engagement research: the individual-level bias of existing measures, the gap between abstract theory and classroom-observable signals, and the difficulties of achieving scalability and equity in authentic learning environments. These challenges motivate the search for interpretable, low-cost proxies that can bridge theory with practice and provide AI-ready inputs for real-time feedback.

### 2.1 From Individual Indicators to Interaction-Informed Engagement

In the learning sciences, engagement has long been conceptualized as a multidimensional construct encompassing behavioral, emotional, and cognitive dimensions (Fredricks et al., 2004). Instruments such as the Student Engagement Instrument (Appleton et al., 2006) consolidate this individual-level focus.

However, once learning becomes collaborative, individual engagement alone cannot explain differences in how groups construct knowledge. Research therefore introduces concepts such as collaborative engagement and socially shared regulation, emphasizing meaning negotiation, feedback, and co-construction rather than surface activity counts (Volet et al., 2009; Hadwin et al., 2018; Järvelä et al., 2023).

Despite conceptual advances, existing measures still rely heavily on surface cues—clicks, time-on-task, or turn counts—which often misrepresent interactional processes (Kuhlman et al., 2024). A cross-level translational pathway is needed to connect mechanisms identified in research with signals that are feasible for real classroom deployment.

## 2.2 Bridging the Neural–Behavioral Gap: From BBS to Observable Proxies

Neuroscience-informed studies show that episodes of shared understanding, adaptive feedback, and consensus often coincide with neural coupling or synchrony during interaction (Stephens et al., 2010; Pan et al., 2018, 2020, 2023). These findings highlight which interaction events matter cognitively, even though such studies rely on EEG/fNIRS hyperscanning and are not scalable for classrooms. Importantly, synchrony research consistently identifies understanding- and consensus-related events as key triggers of cognitively aligned interaction.

Despite these insights, BBS studies face practical barriers. Empirical work shows that higher BBS levels predict more effective problem solving (Du et al., 2022), but such studies depend on costly EEG devices and highly controlled conditions. Multimodal extensions (Lawrence & Weinberger, 2022) offer richer datasets, yet remain complex and intrusive. Technical challenges in data synchronization and interpretation further restrict classroom use (Nasir et al., 2022). Even when neural–behavioral coupling is observed, its benefits materialize only when cognitive gains outweigh attentional costs (Gu et al., 2024). These constraints illustrate the disconnect between neuroscience findings and scalable classroom application.

To address this gap, prior work has proposed a two-phase strategy: calibrating behavioral proxies with EEG to establish cognitive fidelity, then transitioning to behavior-only deployment for scalability (Goodwin, 2022; Pan et al., 2023; Markiewicz et al., 2024). Building on this rationale, the present study operationalizes understanding-related behaviors (Q2)—such as retelling, explaining, or integrating peers' ideas—and consensus behaviors (Q4)—such as agreement building and joint decision making—as mechanism-informed, classroom-observable proxies inspired by synchrony research. While feedback (Q3) is recognized as a regulatory component in neuroscientific work, this framework emphasizes Q2 and Q4 as the core signals that can be reliably externalized in classroom discourse. Rather than treating these behaviors as surface signals, this framework conceptualizes Q2 and Q4 as observable expressions of cognitively meaningful interaction events.

These proxies could be automated in future work through transcription or lightweight NLP, offering a pathway toward scalable analysis without requiring specialized sensing. In this way, neuroscience provides a mechanism-based rationale for designing low-cost, classroom-ready signals for technology-enhanced learning environments. Q2 and Q4 together form the behavioral foundation for the Composite Signal Index (CSI), which models individual engagement within collaborative tasks and supports diagnostic and feedback functions that can strengthen learners' motivation and participation over time.

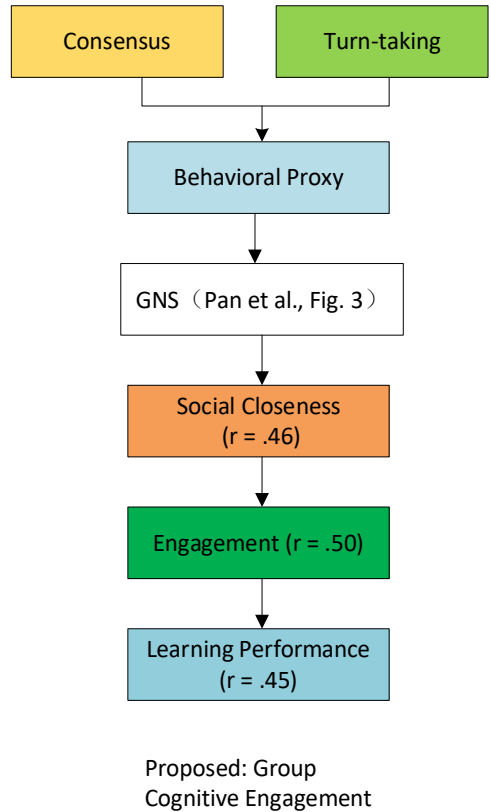

Figure 2.1 Pathway from critical interaction events to classroom-observable engagement signals

Figure 2.1 summarizes a mechanism pathway reported in prior fNIRS hyperscanning studies (Pan et al., 2023), which examined neural synchrony during consensus building. In this study, such findings provide conceptual grounding for translating consensus- and understanding-related events into low-cost, classroom-ready signals for technology-enhanced learning. While the neural work operates at the group level, our empirical framework models individual engagement using behavioral indicators (CSI × Q6).

## 2.3 Scalability and Equity: Limits of Multimodal Sensing and Surface Models

One common approach to measuring engagement relies on surface-level individual behaviors, such as speaking turns, clicks, or time-on-task (Fredricks et al., 2004; Henrie et al., 2015). While these indicators are convenient and low-cost, they are easily misinterpreted: silence does not necessarily mean disengagement, and frequent talk does not guarantee cognitive depth (Kuhlman et al., 2024; Panadero et al., 2023). When extended to groups, these proxies risk overlooking silent or marginalized learners, raising broader concerns of fairness and visibility in AI-supported classrooms. Prior work has warned that AI systems may amplify inequities if they rely on biased indicators (Holstein & Doroudi, 2021), while recent reviews highlight persistent challenges of fairness and bias in educational AI applications (Chinta et al., 2024). Emerging classroom studies further show that learners themselves perceive AI integration as a double-edged sword, valuing efficiency but fearing the erosion of relational and cultural dimensions of teaching (Almashour et al., 2025). Together, these insights underscore that engagement models must be not only scalable and interpretable but also attentive to issues of equity and inclusivity.

A second stream of research has explored multimodal learning analytics (MMLA), which integrate signals such as eye-tracking, facial expressions, prosody, or physiological data (Bosch et al., 2016; D'Mello et al., 2017). These methods can capture fine-grained dynamics and improve predictive precision, but they face practical barriers. Devices are expensive and intrusive, and findings often depend heavily on specific tasks and contexts (Du et al., 2022; Nasir et al., 2022). Large-scale reviews confirm that synchronization, interpretability, and transferability remain major obstacles for real-world adoption (Guerrero-Sosa et al., 2025).

Building on these multimodal methods, recent studies have reported impressive predictive accuracies by combining multiple physiological and behavioral channels. Gupta et al. (2024), for instance, integrated EEG and facial cues to achieve attentiveness recognition rates above 90%, while Jamil and Belkacem (2024) fused EEG with eye-tracking in a CNN model, yielding an AUC of 0.98. These approaches demonstrate the potential of multimodal sensing but also underscore its reliance on complex hardware, intensive computation, and controlled conditions, which limits scalability and ecological validity in authentic classrooms.

Even when such systems achieve predictive accuracy, they often struggle to generalize across contexts. Xing et al. (2022) reported that behavioral and cognitive engagement predicted collaborative problem-solving, whereas social engagement showed negative effects. Hildebrandt and Mehnen (2024) further found that only certain interaction patterns transfer reliably across courses, while most remain highly context-bound.

These findings underscore a critical need: engagement models must be not only cost-effective and scalable but also sensitive to fairness and transferable across diverse classrooms. Lightweight, theory-informed behavioral proxies—designed to reduce misclassification and support real-time instructional use—represent one promising direction.

## 2.4 Translating Lab Insights into Classroom-Ready Design and the CSI × Q6 Framework

Laboratory studies have demonstrated that dual-brain stimulation using transcranial alternating current stimulation (tACS) can modulate neural synchrony and establish causal links between synchrony, behavior, and learning outcomes (Pan et al., 2021). Complementary reviews further show that neurophysiological approaches provide valuable insights into team performance and coordination (Algumaei et al., 2023). Yet these methods remain difficult to apply in authentic classrooms because they require controlled environments, are costly, and lack ecological validity.

An alternative pathway is to embed structured interaction tasks directly into classroom practice. In the present study, collaborative speaking activities were designed to elicit high-engagement signals under natural conditions. The sequence—rotating reading, retelling, feedback, and consensus building—offers repeated opportunities for learners to articulate understanding and negotiate agreement. Unlike laboratory stimulation, this design requires no special equipment, integrates seamlessly with instruction, and can be transferred across subjects and formats. These elicited behaviors are interpretable and can be detected using transcription, video coding, or natural language processing, making them suitable for AI-ready applications. A detailed comparison between laboratory-based stimulation and classroom-induced engagement is provided in Table 3.1 (see Section 3).

Building on this classroom-oriented design, the present study consolidates understanding behaviors (Q2) and consensus behaviors (Q4) into a Composite Signal Index (CSI), and incorporates sustained motivation (Q6) to form the CSI × Q6 quadrant framework. This framework is diagnostic rather than predictive: it is intended to help teachers recognize interaction-informed engagement patterns and to provide transparent, reproducible signals for AI-assisted feedback in technology-enhanced learning environments. By being interpretable, portable, and deployable, the CSI × Q6 framework positions engagement modeling for real classroom use. This approach improves interpretability and avoids the cost–fidelity trade-off of neural sensing, making engagement modeling feasible in real classrooms.

## 3 Research Design: From Neuroscience Inspiration to Structured Group Speaking Tasks

### 3.1 Research Logic and Educational Rationale

This study is positioned as a design-based exploratory validation, not a controlled experiment. Following a design–test–reflection cycle common in educational technology R&D, we designed a structured collaborative speaking task, tested its feasibility in an authentic ESL classroom, and reflected on findings to inform future iterations (e.g., NLP-based automation and rule refinement). The goal is diagnostic rather than predictive: to produce interpretable, classroom-embedded signals that teachers and AI systems can use for timely support. The study followed a design–test–reflection cycle (see Figure 3.1).

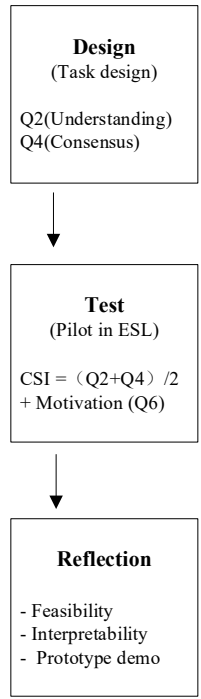

Figure 3.1 illustrates the design–test–reflection cycle that guided the study

Table 3.1 summarizes the educational needs at learning, teaching, and system levels, and illustrates the study's responses.

Table 3.1 Educational Needs and Research Responses

| Level | Educational Need | Research Response |
|---|---|---|
| Learning | Surface indicators (e.g., talk frequency, participation counts) often misclassify engagement: silent students ≠ disengaged; frequent talk ≠ deep thinking. | Structured speaking tasks designed to elicit shared understanding (Q2) and consensus (Q4) as meaningful engagement proxies. |
| Teaching | Teachers lack real-time, interpretable tools to diagnose group engagement. | CSI × Q6 quadrant framework provides a transparent diagnostic scheme; a prototype demonstrates how differentiated feedback can be generated. |
| System | Existing AI tools are often individual-focused, opaque, and difficult to explain. | Introduction of group-level, interpretable proxies that are lightweight, explainable, and scalable across platforms. |

Table 3.1 summarizes the educational needs at learning, teaching, and system levels, and illustrates how the present study responds through the design of interpretable, group-level engagement proxies.

**Educational need**. Existing classroom indicators (e.g., participation counts, speaking duration) are surface-level and often misclassify engagement: frequent talk ≠ deep thinking, silence ≠ disengagement (Thurn et al., 2023). Even behavior-only prediction models can confuse deep concentration with low engagement and over-value high-frequency talk (Apicella et al., 2022; Karimah & Hasegawa, 2022). Meanwhile, multimodal solutions (EEG, eye-tracking, facial analysis) achieve high accuracy in lab-like settings but remain costly and impractical for everyday classrooms. This gap motivates lightweight, interpretable, and scalable proxies that can be embedded into routine instruction.

To address this gap, the present study designed a lightweight collaborative task that translates abstract constructs into classroom-observable signals.

**Design choice**. Building on evidence that shared understanding (Q2) and consensus (Q4) coincide with heightened group alignment and predict collaborative outcomes, we embedded a three-phase task—rotating reading → individual retelling → group consensus—into regular lessons to naturally elicit these signals. The resulting indicators are portable (survey-based, rubric-ready), reproducible (transparent rules), and AI-ready (mappable to dashboards and prompt libraries). The Composite Signal Index (CSI) consolidates Q2 and Q4; combined with Q6 (sustained motivation), it yields the CSI × Q6 quadrant framework for teacher-friendly diagnosis and targeted feedback.

**Laboratory vs. Classroom Approaches**. Laboratory studies have shown that dual-brain stimulation with tACS can modulate neural synchrony and establish causal links between synchrony, behavior, and learning outcomes (Pan et al., 2021). Yet such methods are costly, require controlled environments, and remain far from classroom practice (Algumaei et al., 2023). In contrast, the present study embedded a collaborative speaking task—rotating reading, individual retelling, and group consensus—directly into classroom instruction. This lightweight design eliminated reliance on specialized equipment, enhanced ecological validity, and yielded signals that are interpretable and AI-ready, enabling direct integration into real-time monitoring and feedback systems (see Table 3.2).

Table 3.2 Comparison of Laboratory-Based Neural Stimulation and Classroom-Induced Engagement

| Item | Laboratory Neural Stimulation (e.g., tACS) | Classroom-Induced Engagement (this study) |
|---|---|---|
| Method | Dual-brain stimulation with tACS | Structured three-person collaborative speaking task |
| Context | Controlled, artificial paired tasks | Seamlessly integrated into classroom instruction |
| Induction | External electrical stimulation | Rotating reading + retelling + consensus discussion |
| Causal Chain | tACS → neural synchrony → behavioral synchrony → learning outcome | Natural interaction → shared understanding + consensus → group engagement → learning outcome |
| Limitations / Advantages | High equipment cost, low ecological validity, poor scalability | No device dependency, classroom-embedded, transferable, AI-ready |

As shown in Table 3.2, laboratory-based approaches offer causal insights but are costly, constrained, and far removed from classroom practice. In contrast, the classroom-based design avoids device dependency, enhances ecological validity, and yields behavioral signals that are both interpretable and AI-ready. This positions the framework as a practical response to the educational demand for scalable, real-time, and teacher-relevant engagement signals.

NLP and discourse-based approaches have further demonstrated that collaborative processes can be captured through multiple dimensions of interaction, including peer feedback, I–we positioning, and dialogue dynamics (Castro et al., 2023; Peltoniemi et al., 2025; Park et al., 2025). This growing body of work underscores the feasibility of linking linguistic traces of interaction to diagnostic models of group engagement, providing a foundation for future automation of the CSI framework.

## 3.2 Task Design and Participants

The study took place in a weekend adult ESL program in the San Francisco Bay Area. Nine adult ESL learners enrolled in a weekend TOEFL preparation class participated. Learners represented diverse cultural backgrounds and were assigned to three groups of three to encourage cross-cultural interaction.

The instructional design consisted of a collaborative speaking task involving rotating reading, individual retelling, and group consensus discussion.

In the reading phase, each learner took turns reading aloud from short passages, which ensured equal exposure and sustained attention to content. In the retelling phase, learners summarized the text in their own words, making comprehension visible and thereby generating Q2 signals. Finally, in the consensus phase, learners discussed guiding questions and produced a shared conclusion, which elicited Q4 signals.

The design was deliberately lightweight, required no special equipment, and could be embedded into regular class instruction. Its purpose was not to simulate a laboratory experiment but to demonstrate how everyday classroom activities can elicit meaningful engagement signals.

Table 3.3 Design of the Collaborative Speaking Task and Corresponding Engagement Signals

| Task Phase | Primary Objective | Relation to Engagement Signals | Implementation Details |
|---|---|---|---|
| Rotating Reading | Ensure equal exposure and attention to text | Precondition for Q2 (shared understanding) | Each student reads one sentence; peers support pronunciation/explanation |
| Individual Retelling | Make comprehension explicit | Directly elicits Q2 (shared understanding) | Learners retell main points; brief peer feedback allowed |
| Peer Feedback | Detect divergence and reinforce agreement | Supports transition from Q2 to Q4 | Peers provide clarifications or corrections before consensus |
| Group Consensus | Integrate perspectives, finalize agreement | Directly elicits Q4 (consensus formation) | Group discusses guiding questions; produces shared written conclusion |

Two rounds of tasks were conducted; in the second round, explicit peer feedback requirements were added to strengthen the transition from shared understanding (Q2) to consensus (Q4). Detailed task materials and procedural instructions are provided in Appendix A. Prior neuroscience research has also shown that understanding and consensus behaviors are associated with stronger neural synchrony and improved learning outcomes, which further justified embedding these phases into the task design.

## 3.3 Constructing Behavioral Proxies and the CSI

Signals were captured through structured self-report items and classroom observation. Q2 (understanding) and Q4 (consensus) were retained as the structural basis of the Composite Signal Index (CSI). It remains an extension target for future studies, particularly once NLP-based recognition becomes feasible. Although Q3 (feedback responsiveness) did not enter the CSI computation in this dataset due to limited predictive strength, its linguistic clarity suggests potential for future automation when transcription-based analysis becomes feasible.

The CSI was computed as the average of Q2 and Q4 scores:

$$CSI = \frac{Q2 + Q4}{2}$$

This consolidation provides not only theoretical grounding—as both signals are closely linked with heightened group neural synchrony and collaborative learning outcomes (Pan et al., 2023; Stephens et al., 2010; Zheng et al., 2023)—but also practical advantages in quantification and transferability. The CSI thus establishes a structured and interpretable proxy for group-level engagement and lays the foundation for subsequent predictive modeling and AI-ready feedback systems. Furthermore, Q2, Q3, and Q4 map onto different dimensions of engagement—cognitive, metacognitive, and behavioral/cognitive, respectively (Zheng et al., 2023). This multidimensional mapping suggests that CSI has the potential to serve as a composite proxy integrating across levels.

While Q2 and Q4 formed the CSI core in this dataset, Q3 (feedback responsiveness) was also embedded in the tasks and is summarized here as a candidate proxy. Although excluded from the present CSI calculations due to limited predictive strength, it remains valuable for understanding regulatory processes in dialogue and is preserved as an expansion variable for future research.

Table 3.4 summarizes the candidate behavioral proxies with their cognitive functions, illustrative utterances, and empirical support, highlighting their potential detectability in classroom discourse.

Table 3.4 Candidate behavioral proxies and their empirical support

| Behavioral Proxy | Cognitive Function | Observable Language Examples | Empirical Support |
|---|---|---|---|
| Understanding (Q2) | Semantic processing & predictive processing | "Do you mean…?", "I understand…" | Stephens et al. (2010) |
| Feedback (Q3) | Regulation & conversational progression | "You could… maybe…", "I suggest…" | Markiewicz et al. (2024); Castro et al. (2023) |
| Consensus (Q4) | Group decision-making & perspective integration | "We all think…", "We agree that…" | Pan et al. (2023); Peltoniemi et al. (2025) |

Note. Q3 (Feedback) was not included in the current CSI analysis but is retained here as a candidate proxy, with potential value for future automated modeling."

As illustrated in Figure 3.2, this design extracted Q2 (understanding) and Q4 (consensus) from structured classroom tasks, combined them into the Composite Signal Index (CSI), and examined their association with sustained motivation (Q6). This operational pathway provided a practical basis for testing the viability of CSI and the CSI × Q6 quadrant framework as AI-ready engagement signals.

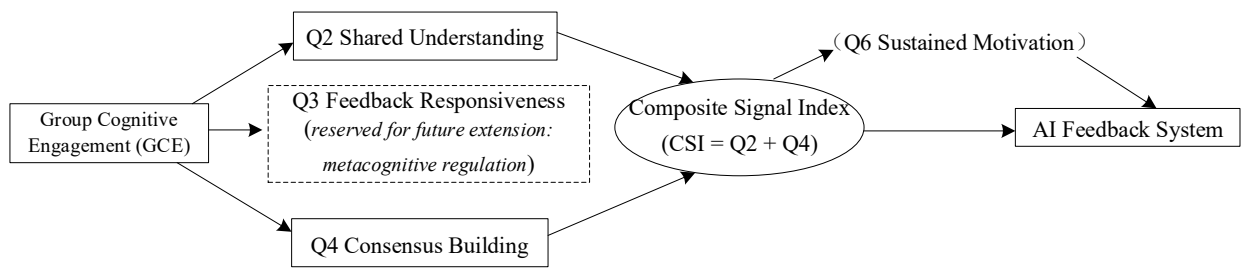

Figure 3.2 Operational pathway of structured signals and the Composite Signal Index (CSI)

## 3.4 Feasibility and Transferability

The method was successfully implemented in a low-resource ESL classroom, which confirmed its feasibility under everyday teaching conditions. The three-phase collaborative speaking task—rotating reading, individual retelling, and group consensus—was simple to administer, promoted balanced participation, and naturally elicited signals of shared understanding and consensus. Beyond the pilot site, the design demonstrates strong transferability. Because the task structure is generic and does not rely on specific textbooks or specialized equipment, it can be adapted to online platforms such as Zoom with transcription-based analysis, and extended to STEM discussions, project-based learning, or seminar-style courses. Group size is also flexible, ranging from dyads to groups of five. Importantly, the CSI × Q6 framework and its accompanying prototype are positioned as proof-of-concept demonstrations. They are intended to illustrate feasibility and educational potential rather than to claim system-level deployment. This distinction situates the study within the scope of design-based educational technology research, which provides early evidence of feasibility, identifies areas for refinement, and lays the foundation for subsequent iterations.

## 3.5 Diagnostic Framework

To enable interpretable diagnosis of group engagement, the Composite Signal Index (CSI) was combined with sustained motivation (Q6) to form a two-dimensional quadrant framework. This framework distinguishes four engagement states: High CSI–High Q6 (ideal), High CSI–Low Q6 (motivationally insufficient), Low CSI–High Q6 (structurally insufficient), and Low CSI–Low Q6 (at-risk). The goal of this scheme is not statistical classification, but rather to provide teachers with a transparent and practice-oriented tool to support differentiated instructional feedback. As an exploratory supplement, a decision-tree representation was also considered to visualize diagnostic rules and potential threshold effects. It is important to emphasize that the decision tree serves only as a complementary tool to enhance interpretability and usability, and not as a core method. Detailed analyses and examples of this exploratory representation are presented in Section 4.5. At the methodological level, it is sufficient to note that while the decision tree illustrates potential extensions, the core diagnostic logic remains the CSI × Q6 quadrant model.

## 3.6 Data Collection, Analysis, and Reproducibility

The study was conducted in a weekend TOEFL preparation class at a language school in California, involving nine adult ESL learners who were randomly assigned to three groups of three. Each group completed two rounds of the collaborative speaking task, which included rotating reading to ensure equal participation and shared attention, individual retelling to externalize comprehension processes, and group consensus discussion to reach agreement on guiding questions and produce a written outcome. Immediately after each round, learners completed a short self-evaluation survey on a five-point Likert scale. The key items captured willingness to speak (Q1), shared understanding (Q2), feedback responsiveness (Q3), consensus formation (Q4), confidence in sharing opinions (Q5), and sustained motivation (Q6). The survey also included an open-ended question inviting students to reflect on which part of the activity was most helpful.

For the quantitative analysis, descriptive and exploratory modeling procedures were applied. These included correlation and regression analyses to examine the relationship between behavioral signals and sustained motivation, the construction of the Composite Signal Index (CSI) as the average of Q2 and Q4, quadrant classification combining CSI and Q6 to differentiate engagement states, and exploratory decision-tree modeling to visualize threshold effects and diagnostic pathways. The qualitative analysis involved thematic mapping of open-ended responses to identify references to understanding, feedback, and consensus, which served to triangulate with the quantitative findings.

The study also emphasized transparency and reproducibility. The three-phase speaking task does not require specialized equipment and can be applied in any collaborative learning context, including STEM discussions and project-based learning. Survey items targeting shared understanding and consensus can be reused across disciplines wherever collaboration requires comprehension and joint decision-making. The CSI formula and quadrant classification rules are transparent and can be replicated easily from survey data. Looking ahead, the diagnostic prompts illustrated in Tables 4.4 and 4.4a can be further standardized into reusable templates for teachers or AI systems to generate consistent instructional feedback.

## 4 Results and modelling analyses

### 4.1 Signal selection and correlation analyses

We begin by reporting bivariate associations and simple univariate regressions for Q1–Q6 (see Figures 4.1–4.2 and Table 4.1). All analyses reported below are exploratory and based on the pilot sample (N = 9, three 3-person groups); results should therefore be interpreted with caution.

Pearson correlations and univariate regressions indicated that Q1 (willingness to speak / immediate expression) had the largest observed association with sustained motivation (Q6) in this dataset (r = 0.77; $R^2$ ≈ 0.60; p < .05), suggesting that surface expression relates to concurrent reports of motivation in this sample. Q2 (shared understanding) and Q4 (consensus building) showed moderate positive associations with Q6 ($R^2$ ≈ 0.30–0.36 across univariate models), which is consistent with theoretical accounts that link cognitive processing and social coordination to group engagement.

By contrast, Q3 (feedback responsiveness) and Q5 (self-confidence) explained little variance in Q6 in these univariate models and were therefore not retained for subsequent modeling in this pilot study. We retained Q2 and Q4 as the structural signals for further analysis on the basis of (a) their empirical relationships with Q6 in this sample, and (b) their theoretical relevance and classroom observability. Although Q3 did not contribute predictive utility here, prior work highlights feedback's potential role in supporting interaction and cognition (Markiewicz et al., 2024; Castro et al., 2023); Q3 therefore remains a candidate signal for future, larger-scale work (particularly where automated detection is available).

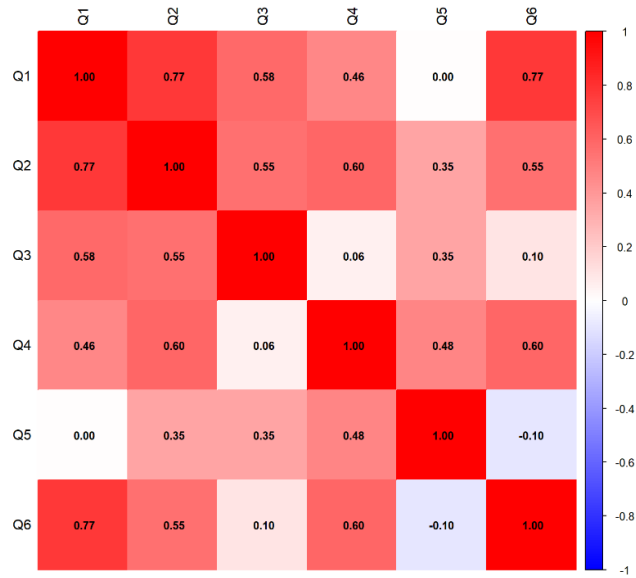

Figure 4.1 Correlation heatmap among Q1–Q5 and Q6

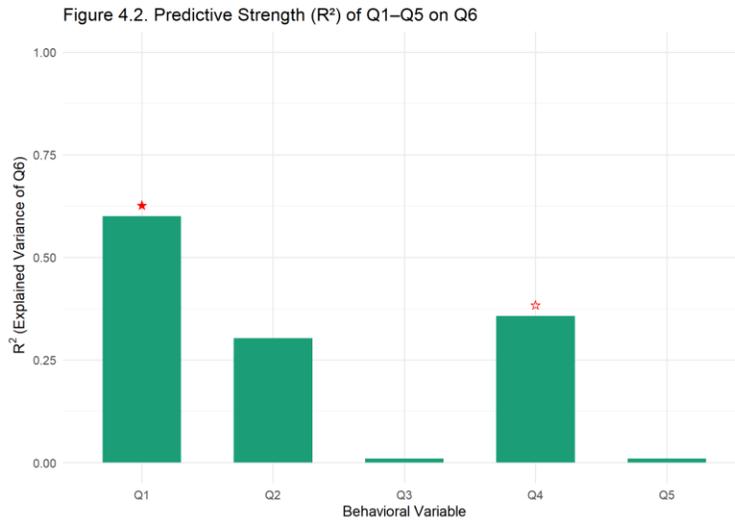

Figure 4.2 Comparative explanatory power (R²) of Q1–Q5 for Q6 (univariate models)

Table 4.1 Univariate regression results: Predictive ability of Q1–Q5 for Q6 (sustained motivation)

| Predictor | Coefficient (β) | Std. Error | p-value | R² | Short interpretation |
|-----------|-----------------|------------|---------|-----|----------------------|
| Q1 (Immediate expression) | 0.333 | 0.103 | 0.014 | 0.600 | Largest observed association in this sample; preliminary evidence of a motivational output signal |
| Q2 (Shared understanding) | 0.550 | 0.316 | 0.125 | 0.303 | Positive, exploratory relationship; retained as a structural signal |
| Q3 (Feedback responsiveness) | 0.100 | 0.376 | 0.798 | 0.010 | Little explanatory power in this dataset; candidate for future work |
| Q4 (Consensus building) | 0.714 | 0.362 | 0.089 | 0.357 | Suggestive association (exploratory); retained as a structural signal |
| Q5 (Self-confidence) | −0.100 | 0.376 | 0.798 | 0.010 | Little explanatory power; not retained in subsequent modeling |

Note. Exploratory univariate regression results; see Methods 3.2 for details.

Statistical analyses and visualizations were performed in R (version 4.4.3).

## 4.2 CSI and its association with sustained motivation

### 4.2.1 Regression analyses of CSI and Q6

We combined Q2 and Q4 into a Composite Signal Index (CSI) representing the structural dimension of group engagement (see Methods for the exact calculation). A simple OLS regression of Q6 on CSI yielded moderate explanatory power in this pilot sample ($R^2 \approx 0.41$). The estimated coefficient for CSI was positive ($\beta = 0.78$, SE = 0.35), and the associated p-value was 0.064. Given the small sample size, this result should be read as preliminary evidence that higher CSI values tend to be associated with higher sustained motivation; the finding is indicative rather than confirmatory.

Table 4.2. Regression analysis showing the association between CSI and sustained motivation (Q6)

| Parameter | Value |
|-----------|-------|
| Intercept | 0.926 |
| Coefficient (β) | 0.778 |
| Standard Error (SE) | 0.354 |
| t-value | 2.198 |
| p-value | 0.064 |
| R² | 0.408 |
| Adjusted R² | 0.324 |
| F-statistic | 4.831 |
| Degrees of freedom | 1, 7 |
| Residual standard error | 0.433 |

Note. Results are based on exploratory regression analyses implemented in R (N = 9). The reported p-value (p = 0.064) indicates marginal evidence under conventional thresholds; larger samples are required for confirmatory inference.

### 4.2.2 Quadrant classification for instructional differentiation

Drawing on the complementary roles of Q2 (shared understanding) and Q4 (consensus), we combined the Composite Signal Index (CSI) with Q6 to form a two-dimensional quadrant framework (Figure 4.3). Rather than serving as a statistical classifier, this framework is designed as a transparent and practice-oriented diagnostic tool to help teachers and AI systems differentiate patterns of engagement and provide targeted feedback.

The framework distinguishes four engagement profiles:

1. High CSI – High Q6 (Ideal state): Strong structural signals and high reported motivation, reflecting an optimal collaborative engagement state that can be sustained and modeled.

2. High CSI – Low Q6 (Motivational deficit): Structural understanding and consensus are solid, but motivation is low (e.g., fatigue or decreased affect). Teachers may focus on motivational prompts while maintaining structural strengths.

3. Low CSI – High Q6 (Structural deficit): Motivation is present but markers of shared understanding and consensus are weak. This suggests activity without shared cognitive coordination, calling for prompts to strengthen perspective-taking and alignment.

4. Low CSI – Low Q6 (At-risk state): Both structural and motivational indicators are low, signaling disengagement risk and the need for adaptive re-engagement strategies.

The thresholds for "high" versus "low" in this pilot were derived from the observed sample distribution and are presented illustratively. In classroom use, these thresholds can be flexibly calibrated to local contexts or larger datasets.

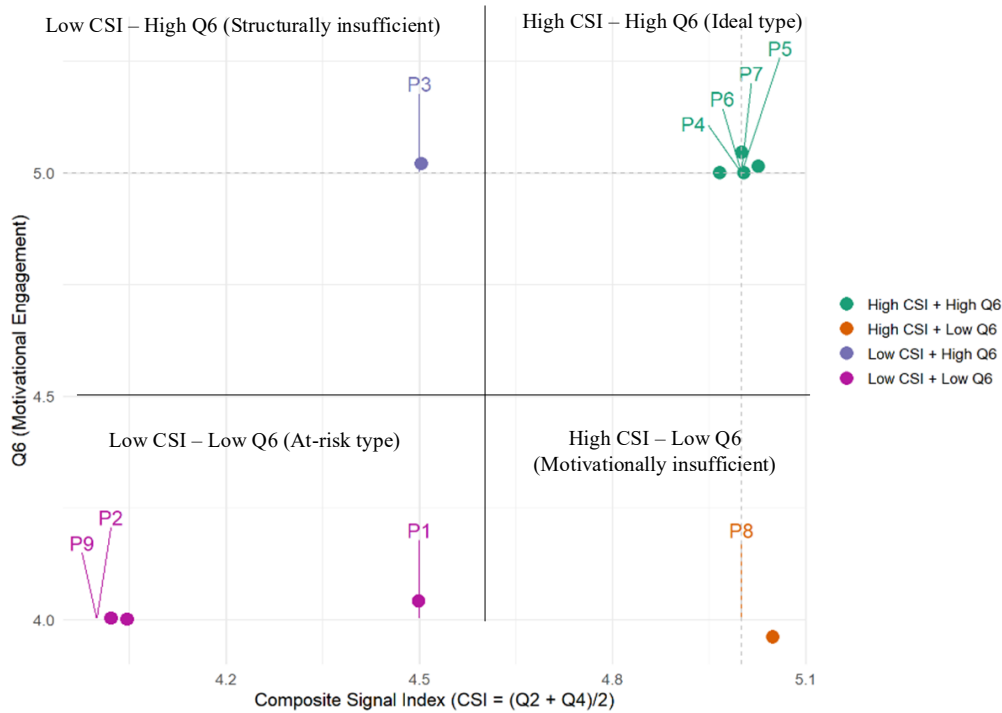

Figure 4.3 CSI × Q6 quadrant framework (illustrative thresholds based on sample distribution)

## 4.3 Open-Ended Responses

In addition to quantitative analyses, open-ended responses from students provided qualitative evidence that further validated the signal framework. Learners' reflections consistently highlighted behaviors aligned with shared understanding (Q2), consensus formation (Q4), and feedback (Q3), reinforcing the ecological validity of the proposed proxies.

Several students explicitly mentioned that "sharing information in group" and "discussing the text" helped them reach agreement, directly corresponding to Q4 (consensus). Others emphasized that they "understood peers' ideas" or benefited from "reading and listening to other people's opinions", reflecting Q2 (shared understanding). Still others highlighted the role of feedback in sustaining learning ("speaking activity," "discussion activities helped me a lot"), underscoring feedback's function as a bridge between understanding and consensus.

Qualitative evidence aligned with quantitative results in reinforcing the validity of the signal framework. Students' reflections echoed the identified proxies of shared understanding (Q2), feedback (Q3), and consensus (Q4), and demonstrated that these signals are recognizable in natural classroom discourse. This alignment strengthens the ecological validity of the CSI pathway and highlights its diagnostic value for modeling group engagement.

Table 4.3 summarizes representative student quotes, their mapping to engagement signals, and the corresponding interpretation. These results indicate that learners themselves perceived understanding, consensus, and feedback as meaningful drivers of engagement, providing qualitative support for the CSI × Q6 framework.

Table 4.3 Mapping of Open-Ended Responses to Engagement Signals

| Student Quote | Engagement Signal (Q1–Q6) | Interpretation |
|---|---|---|
| "Share what I was understanding about the text." | Q2 – Understanding | Reflects individual comprehension made explicit to peers, consistent with shared understanding. |
| "Reading and listening to other people's opinions." | Q2 – Understanding | Indicates comprehension of peers' input and alignment around meaning. |
| "We had to discuss the text. It was so good." | Q4 – Consensus Formation | Highlights group discussion as a mechanism to reach agreement. |
| "Sharing information in group." | Q4 – Consensus Formation | Information exchange viewed as part of the process of reaching group consensus. |
| "Reading & discussion activities helped me a lot." | Q3 – Feedback / (Q2 – Understanding) | Matches Q3 item on feedback improving ideas; also implies comprehension gains. |
| "The idea of new theory… we can select our own instead of natural selection." | Q4 – Consensus Formation | Emphasizes making a collective choice, aligning with the Q4 construct. |
| "Speaking activity" | Q1 – Speaking Willingness | General positive evaluation of speaking, linked to willingness to participate verbally. |
| "Thank you" | Affective engagement (outside Q1–Q6) | Expression of appreciation and positive emotion; not directly tied to structural signals. |

Table 4.3 summarizes representative student quotes, their mapping to engagement signals, and the corresponding interpretation. These results indicate that learners themselves perceived understanding, consensus, and feedback as meaningful drivers of engagement, providing qualitative support for the CSI × Q6 framework.

Q2 (shared understanding), Q3 (feedback responsiveness), and Q4 (consensus formation) thus emerge not only as theoretically grounded constructs but also as systematically identifiable signals in classroom discourse. This dual role highlights their potential to be translated into teacher-facing dashboards and AI-ready feedback systems, providing a direct bridge from verbal expressions to diagnostic feedback.

## 4.4 AI-Ready Prototype Demonstration

Building on these observable signals, this section presents a prototype as a proof-of-concept rather than an empirical result. The design illustrates how the CSI × Q6 framework can be visualized in a teacher-facing dashboard and mapped to differentiated prompts. Following a simple Signal → Decision → Intervention logic, the prototype demonstrates how questionnaire-derived Q2 (shared understanding) and Q4 (consensus) can be aggregated into the Composite Signal Index (CSI), combined with Q6 for state classification, and then translated into actionable feedback for individuals and groups. Figure 4.4 provides an illustrative example for Group P2 classified as High CSI–Low Q6, showing how such classifications can be mapped onto simple pedagogical moves.

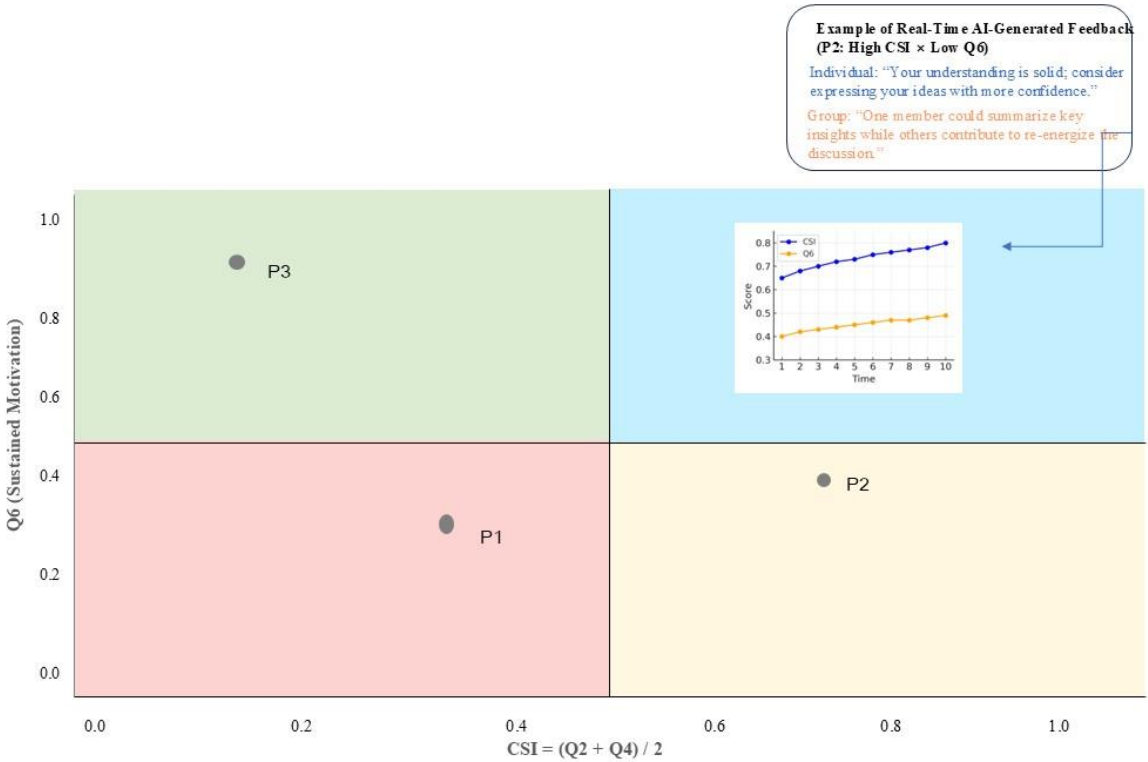

Figure 4.4 Prototype design for group engagement classification and AI-driven feedback (Group P2 example)

The prototype highlights the proof-of-concept function of the framework. It demonstrates how compact, explainable signals can be visualized and linked to actionable suggestions, while avoiding the claim of a validated automated tutor. The design emphasizes transparency and teacher-in-the-loop interaction, consistent with recent work on interpretable machine learning for classroom engagement (Johnston et al., 2024).

Finally, the prototype not only classifies CSI–Q6 states and generates feedback but also illustrates potential explanations for low engagement conditions. As summarized in Table 4.4a, possible causes of the Low CSI–Low Q6 state can be interpreted through multiple theoretical lenses: task difficulty relates to regulatory mechanisms in Social Cognitive Theory (SCT); lack of consensus or meaning reflects Self-Determination Theory (SDT); mismatches between task demands and learner investment are consistent with the Engagement–Fit framework; and the absence of adaptive leader–follower dynamics echoes findings from group neural synchrony research.

## 4.5 Pathways and Diagnostic Value

The purpose of the pathway analysis is to illustrate how AI-ready signals (Q2, Q4, CSI, Q1, and Q6) can be organized into diagnostic logics that inform instructional feedback, rather than to establish causal mechanisms. Exploratory decision-tree models were used as proof-of-concept demonstrations to support diagnostic reasoning and to illustrate how thresholds can be visualized for teachers and AI systems.

The main pathway represents a direct link between the Composite Signal Index (CSI), which consolidates shared understanding (Q2) and consensus (Q4), and sustained motivation (Q6). The decision-tree model (Figure 4.5a) indicated that lower CSI values were consistently associated with reduced willingness to continue, whereas higher CSI values corresponded with stronger motivation. For practice, this pathway suggests a straightforward heuristic: when structural coordination appears weak, learners are at greater motivational risk. Teachers could respond with encouragement or scaffolds that sustain participation.

The reasoning pathway introduces a layered diagnostic lens. In this pathway, Q2 and Q4 were associated with willingness to speak (Q1), which in turn was linked to sustained motivation (Q6). The decision-tree visualizations (Figures 4.5b and 4.5c) showed that when understanding or consensus was low, learners often hesitated to express themselves, and this hesitation predicted lower sustained engagement. For instructional use, this reasoning path highlights the role of interventions such as structured turn-taking, paraphrasing exercises, or consensus checks to strengthen immediate expression before motivational decline occurs.

①Main Path (CSI → Q6)

The decision tree illustrates the individual-level relationship between the Composite Signal Index (CSI) and sustained motivation (Q6). Each split corresponds to a threshold of CSI, and the leaf nodes report the number of individuals (n) and their mean Q6 scores. Results indicate that individuals with CSI ≥ 4.8 consistently reached the highest levels of sustained motivation, whereas individuals with CSI < 4.3 exhibited notably lower motivation. This "main path" highlights CSI as a reliable individual-level proxy for predicting sustained engagement.

**Decision Tree: CSI → Q6**

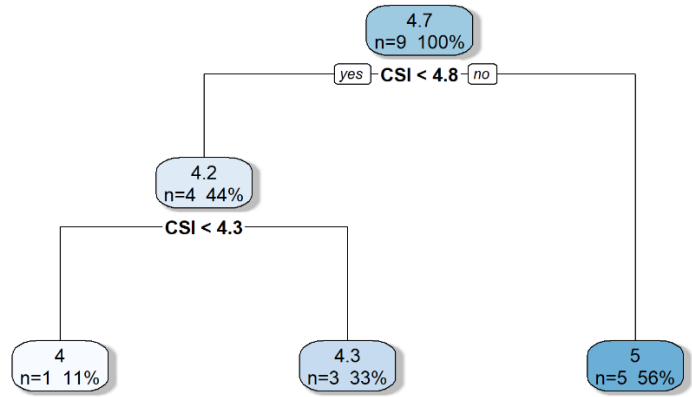

Figure 4.5a Decision tree (Main Path: CSI → Q6)

②Reasoning Path (Q2, Q4 → Q1_Story2 → Q6)

To further explore the mechanism, we divided the analysis into two stages. Figure 4.5b demonstrates how shared understanding (Q2) and consensus (Q4) jointly predicted immediate expression motivation (Q1_Story2). Building on this, Figure 4.5c shows that Q1_Story2 in turn predicted sustained motivation (Q6), again with threshold effects indicating that higher levels of immediate expression were strongly linked to stronger sustained motivation.

Q1_Story2 refers to students' immediate expression motivation measured after the second story, rather than the average across both tasks.

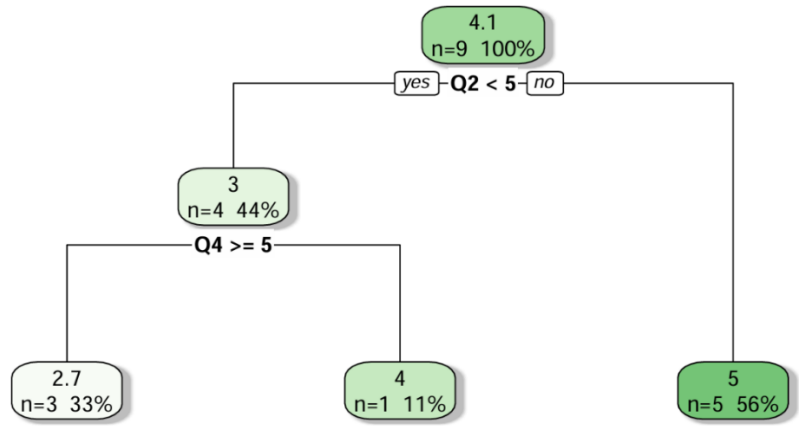

Figure 4.5b Decision tree (Reasoning Path, Stage 1: Q2, Q4 → Q1_Story2)

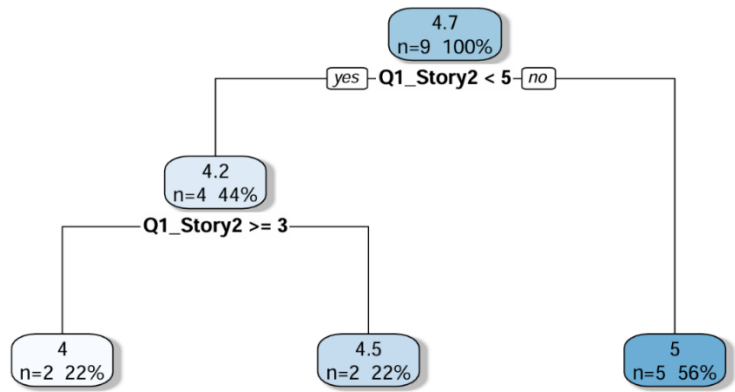

Figure 4.5c Decision tree (Reasoning Path, Stage 2: Q1_Story2 → Q6)

These exploratory models demonstrate how compact behavioral signals can be translated into actionable logics. The main path enables rapid screening for motivational risks, while the reasoning path provides anticipatory insights into specific bottlenecks—whether comprehension, consensus, or expression. In both cases, the contribution lies in offering interpretable rules that can be adapted into explainable, teacher-facing feedback systems.

Table 4.4 Engagement Patterns and Diagnostic Feedback Prompts

| Engagement Pattern | Trigger Condition | Suggested Diagnostic Feedback |
|---|---|---|
| High CSI – High Q6 (Ideal State) | CSI high, Q6 high | Continue with the current level of participation, and try to help peers deepen their understanding. (Well-balanced state, no immediate intervention needed.) |
| High CSI – Low Q6 (Motivational Deficit) | CSI high, Q6 low | Your understanding is solid; try expressing your ideas with more confidence. This state suggests teachers should provide motivational prompts despite strong structural signals. |
| Low CSI – High Q6 (Structural Deficit) | CSI low, Q6 high | Before expressing your views, listen to more peers to make your argument more convincing. (Motivation is present but structural coordination is weak.) |
| Low CSI – Low Q6 (At-Risk State) | CSI low, Q6 low | Let's try a different angle or approach to see if we can find a new entry point. (May result from task difficulty, low morale, dominance effects, or lack of relevance—see Table 4.4a.) |
| Low Q2 (Insufficient Understanding) | Q2 below threshold | You may not be fully sure about your peers' ideas; try paraphrasing their viewpoints in your own words. |
| Low Q4 (Lack of Consensus) | Q4 below threshold | Have you reached a consensus? Try summarizing your current ideas together to check for agreement. |
| Low Q1 (Low Expression Willingness) | Q1 below threshold | Your collaborative attitude is great, but you could actively share your ideas to bring more voices into the discussion. |

Table 4.4 summarizes diagnostic patterns and corresponding feedback prompts derived from the CSI–Q6 quadrant framework and reasoning pathways. The first four patterns represent quadrant-based classifications for rapid judgment, while the latter three highlight specific bottlenecks (understanding, consensus, or expression willingness). The "Low CSI – Low Q6" state may reflect multiple underlying causes, which are elaborated in Table 4.4a.

Table 4.4a. Possible Causes of the "Low CSI – Low Q6" State and Corresponding Teacher/AI Responses

| Possible Cause | Diagnostic Feature | Suggested Teacher/AI Response |
|---|---|---|
| Task too difficult | Students confused, unable to retell or reach consensus | Simplify task, provide scaffolding materials, clarify instructions |
| Group morale/emotional low state | Group appears passive, low energy, little initiative | Use ice-breakers, short energizing activity, or emotional check-in |
| Dominance or marginalization | One or two students dominate while others remain silent | Enforce turn-taking, assign roles, encourage equal participation |
| Task lacks relevance/meaning | Students show disinterest or treat task as mechanical | Reframe task with real-life examples, connect to learners' goals |
| Lack of adaptive leader–follower dynamics | Everyone tries to adjust simultaneously (overcompensation), or no one adjusts (stagnation) | Encourage one student to summarize or lead temporarily while others follow, then rotate roles to maintain balance |

Table 4.4a further unpacks the Low CSI–Low Q6 state by outlining multiple underlying causes, ranging from task difficulty to group morale and interactional imbalance. Each cause corresponds to distinctive classroom features and suggests targeted teacher or AI responses. This extension emphasizes that quadrant states are not monolithic; rather, they can be differentiated through finer-grained diagnostic cues. Such differentiation provides the basis for adaptive intervention design and opens a pathway toward system-level implementation. In addition to task-related and affective influences, recent work highlights that effective cooperation often relies on flexible leader–follower dynamics, in which one member temporarily takes initiative while others align (Markiewicz et al., 2024).

## 4.6 Application Prospects and System Deployability

The integration of the quadrant framework, prototype, and decision-tree analyses points to a pathway from classroom feasibility toward scalable deployment. Building on the demonstrated feasibility of structured tasks in low-resource classroom conditions, this section considers how engagement models might be embedded into learning management systems (LMSs) or AI-assisted platforms to support real-time monitoring and feedback.

The quadrant framework provides a transparent structure for differentiated support. Groups positioned in the High CSI–Low Q6 quadrant can be identified as having strong structural coordination but reduced motivation, suggesting the need for motivational prompts. Conversely, groups in the Low CSI–High Q6 quadrant may benefit from structural guidance to maintain cognitive alignment and consensus. In this way, the quadrant model offers teachers and systems a straightforward heuristic for tailoring interventions.

The prototype extends this logic by demonstrating a closed-loop process. Signals from Q2 and Q4 are aggregated into CSI, combined with Q6, and visualized as feedback at both individual and group levels. Although conceptual in nature, this demonstration illustrates how interpretable indicators can be mapped onto simple pedagogical moves in a teacher-facing dashboard.

Decision-tree analyses further contribute to the system perspective by illustrating how threshold-based rules can be derived. The main path (CSI → Q6) enables rapid identification of motivational risks, while reasoning pathways (Q2/Q4 → Q1 → Q6) provide more detailed diagnostic insights. These complementary mechanisms highlight the potential for adaptive interventions that address distinct breakdowns, such as insufficient understanding, weak consensus, or low willingness to speak.

Collectively, the quadrant framework, prototype, and decision-tree models illustrate an AI-ready engagement pathway that emphasizes explainability and interpretability. The findings remain exploratory but establish a practical foundation for future integration into scalable, teacher-supportive systems. This transition from proof-of-concept validation to prospective deployment provides the basis for the broader discussion in Section 5.

Composite Signal Index (CSI), and shown to be feasible in authentic classroom settings. Regarding RQ2, exploratory analyses revealed a positive association between CSI and sustained motivation (Q6). While larger-scale studies are needed, the initial evidence suggests that the CSI × Q6 quadrant framework can differentiate engagement states and provide diagnostic logic for targeted teacher interventions. Finally, with respect to RQ3, the teacher-facing prototype and quadrant-based prompts demonstrate that these structured signals are not only transparent and interpretable but can also be embedded into a signal–judgment–feedback loop as AI-ready inputs, supporting real-time teacher–AI collaboration and more equitable recognition of learners.

In sum, the exploratory validation provided by this study offers initial answers to the three research questions and, through the positioning of a "Bridging Stage," lays a foundation for future iterations and expansion.

# 5 Discussion

## 5.1 Theoretical and empirical insights

This study proposed a cross-level translational pathway in which understanding and consensus-related mechanisms—validated in prior EEG/fNIRS hyperscanning studies—were operationalized into classroom-observable signals. The positive association between CSI and sustained motivation (Q6) suggests that structural signals of understanding and consensus can capture motivational dynamics, thereby providing an interpretable diagnostic tool for teacher-led monitoring and AI-assisted classification.

The theoretical contribution of this work does not lie in introducing a new construct, but in demonstrating a mechanism-informed and cost-efficient approach to engagement modeling. By moving beyond surface indicators such as frequency of talk or time on task, the CSI framework emphasizes structural signals that reflect shared cognitive understanding and collaborative cohesion. In doing so, it resonates with Social Cognitive Theory (SCT), Self-Determination Theory (SDT), and perspectives on engagement–context fit, strengthening the explanatory value of the model while maintaining transparency and scalability.

On the empirical side, the study carried out a proof-of-concept validation in an authentic ESL classroom. Findings showed that even in a small-sample setting, structured collaborative tasks could elicit mechanism-aligned engagement features. While Q2 and Q4 formed the core components of CSI, Q3 (feedback responsiveness) was intentionally embedded in the task design as a bridge between understanding and consensus. Although Q3 did not enter the CSI calculation due to its limited predictive strength, classroom observations and learner reflections underscored its educational value. For instance, a simple teacher reminder between two task rounds—"please provide more positive feedback and support each other"—resulted in a visible increase in students' feedback behaviors. This example illustrates that feedback operates as a regulatory mechanism in collaborative learning and highlights its potential as a candidate for future NLP-based recognition.

Overall, the contribution of this study is not a replication of neuroscientific experiments but rather the translation of laboratory findings—such as Pan et al.'s (2023) demonstration that consensus triggers neural synchrony—into classroom-friendly proxy signals. In doing so, the study establishes an AI-ready modeling pathway that is theoretically grounded, empirically feasible, and directly relevant for real-time feedback in collaborative learning contexts.

## 5.2 Educational and practical implications and AI-ready potential

Beyond its theoretical contribution, the study also demonstrates value for teaching practice and the design of AI systems. The CSI × Q6 framework provides teachers with a lightweight and low-cost diagnostic tool. By drawing on the quadrant logic, teachers can quickly distinguish different group states. Groups with high CSI but low Q6 indicate sufficient structure but insufficient motivation, where motivational prompts or task adjustments may help sustain engagement. Conversely, groups with low CSI but high Q6 reveal strong motivation but weak structure, where interventions such as role allocation or guided restatements can strengthen collaboration. This differentiated logic enables teachers to intervene with greater precision in the classroom.

The framework also offers a transparent and interpretable pipeline for AI system design. Unlike black-box models, the quadrant classification and decision-tree logic foreground rules that are both accessible to teachers and replicable across contexts. This not only builds teacher trust but also sketches a blueprint for teacher–AI complementarity, where AI systems detect potential risks and provide prompts, while teachers remain the central agents of judgment and intervention. Platforms can leverage the CSI × Q6 logic to trigger gentle, equity-aware nudges (e.g., regrouping, turn rotation, or pacing adjustments) while keeping teachers in the loop. A teacher-facing dashboard foregrounds indicators such as unreciprocated feedback, turn-taking imbalance, and stalled consensus.

Equity considerations further underscore the importance of the framework. The findings highlight that "silence does not equal disengagement, and frequent talk does not guarantee quality." In collaborative learning, silent students may be engaged in deep thinking yet risk being overlooked, while dominant speakers may appear active without contributing substantively. By capturing signals of shared understanding and consensus, teachers can better identify marginalized learners and ensure that their contributions are visible. This aligns with recent scholarship on the complexity of emotion and motivation in engagement: learners' participation can be enhanced through supportive feedback, but it can also diminish under anxiety or fatigue. Both AI systems and teachers therefore play complementary roles in sustaining balanced forms of engagement.

Finally, the framework is highly transferable. Although it was validated in ESL settings, its design is not limited to language learning. It can be adapted to STEM discussions, project-based learning, flipped classrooms, seminar formats, and online or hybrid environments. Group sizes ranging from dyads to larger teams can adopt the same structure, and any collaborative task that involves cycles of comprehension, restatement, and consensus can be mapped to this framework.

Overall, the CSI × Q6 framework illustrates three areas of practical value: it supports real-time classroom diagnosis and differentiated feedback, it establishes a lightweight and interpretable pipeline for AI-ready engagement monitoring, and it safeguards equity by making diverse learners visible across varied educational contexts.

## 5.3 Complementarity with existing research

Table 5.1 summarizes how this study complements representative strands of prior work.

Table 5.1 Complementarity Between Existing Studies and the Present Study

| Reference | Research Focus / Method | Core Contribution | Difference from Present Study | Contribution of Present Study |
|---|---|---|---|---|
| Pan et al., 2023 | fNIRS hyperscanning; examined GNS during consensus and its relation to outcomes | Identified consensus as a specific trigger of GNS, predictive of learning results | Translated consensus into an observable behavioral proxy (Q4) and combined it with Q2 to form CSI | Provided higher-order collaborative signals that are AI-recognizable and deployable |
| Peltoniemi et al., 2025 | Analyzed CSCL dialogues for I–we consensus expressions | Verified that I–we consensus is a critical marker of successful groups | Combined I–we expressions with Q4 to test predictive power for Q6 | Strengthened the structural value of Q4 signals |
| Castro et al., 2023 | NLP classification of six categories of peer feedback | Demonstrated that cognitive feedback can be identified through linguistic features | Incorporated Q3 feedback behaviors into CSI and mapped them to AI-driven feedback | Provided evidence of Q3's modeling potential |
| Zheng et al., 2023 | Automated group engagement analysis with deep learning; demonstrated feasibility of real-time feedback integration. | Validated the feasibility of automated classification and personalized feedback | Replaced black-box models with a lightweight, interpretable CSI framework | Supported a closed loop of signal detection–state judgment–feedback generation |
| Reinero et al., 2021 | EEG synchrony predicting team performance | Showed neural synchrony as a stronger predictor than self-reports | Migrated the neural synchrony logic into behavioral proxies | Built a pathway from neuroscience insights → behavioral proxies → AI-ready applications |

| Gao et al., 2025 | Constructed ESL discussion corpus; evaluated facilitation strategies | Quantified effects of different facilitation strategies on interaction quality | Linked facilitation effects with CSI modeling | Provided a modular approach for strategy-triggered feedback generation |
| Ding & Yusof, 2025 | AI conversational agents for L2 speaking | Improved fluency, accuracy, and reduced anxiety | Integrated oral feedback mechanisms into collaborative tasks | Verified the cross-domain transferability of oral feedback mechanisms |
| Valdesolo & DeSteno, 2011 | Effects of synchronous movement on cooperation and trust | Demonstrated that synchrony enhances social bonding and cooperative intent | Transferred synchrony effects into task design | Provided theoretical grounding for the social–motivational pathway of the CSI framework |

Prior studies have advanced the modeling of group engagement through diverse pathways, ranging from neural synchrony (Pan et al., 2023; Reinero et al., 2021) to discourse analysis (Peltoniemi et al., 2025) and automated NLP pipelines (Castro et al., 2023; Zheng et al., 2023). Building on these contributions, the present study extends this body of work by translating consensus-related neural mechanisms into observable behavioral proxies (Q4), combining them with shared understanding (Q2) to form the CSI, and linking this index with sustained motivation (Q6). In doing so, it establishes an interpretable and classroom-feasible proxy framework.

Unlike approaches that rely on costly neuroimaging or opaque computational models, this framework emphasizes lightweight design, transparency, and reproducibility. Classroom observations and learner reflections provided additional signs of ecological validity: students reported that "reading and discussion activities were very helpful," while teachers noted deeper comprehension during retelling and consensus phases. These insights suggest that structured oral tasks not only increased language output but also stimulated higher-order reasoning and collaborative thinking, underscoring their pedagogical value.

Accordingly, this study not only complements the three established strands of research—neural synchrony, discourse analysis, and NLP automation—but also contributes a transparent and reproducible proxy system at the classroom level, thereby laying a practical foundation for subsequent AI integration.

## 5.4 Limitations

Several limitations of this study should be acknowledged. First, the sample size was small, involving only nine learners, and the findings should therefore be regarded as exploratory rather than confirmatory. Second, the study lacked strict control conditions, such as tasks with or without explicit prompts for understanding or consensus, meaning that the results are better interpreted as correlational rather than causal. Third, signal recognition relied on questionnaires and manual coding. While this approach ensured transparency and interpretability, it limited scalability; future research will need to integrate automated transcription, NLP, and multimodal analytics to achieve real-time recognition. Finally, the empirical setting was restricted to a small adult ESL classroom, which constrains generalizability. Replications in K–12 environments, STEM subjects, and online or blended contexts will be necessary to examine transferability.

Despite these constraints, the central contribution of the study remains intact: it demonstrates a translational pathway from neuroscience-inspired mechanisms to classroom application. In this sense, the limitations highlight the exploratory and generative value of the research rather than diminishing its significance.

## 5.5 Future research directions

Future research can extend this framework along three lines. First, signals at critical phases of collaborative tasks merit closer attention. Early moments of shared understanding and immediate post-feedback responses may hold strong diagnostic value for subsequent motivation and learning outcomes. Future studies could strengthen the capture of these signals through short phase-end surveys or automated analyses of speech and text. While neuroscience data may occasionally serve as a calibration tool, the long-term goal remains to rely on lightweight behavioral proxies to sustain real-time interventions.

Second, the CSI–Q6 quadrant framework should be expanded and tested in larger samples to examine its applicability at both individual and group levels. A systematic library of quadrant-specific feedback strategies also needs to be developed. By linking quadrant classifications with differentiated feedback intensities—direct, scaffolded, or delayed—AI systems could provide adaptive support while preserving transparency and interpretability for teachers.

Third, attention should be given to cross-context transfer and user experience. Future work should test the framework in STEM subjects, K–12 settings, and online or hybrid environments, as well as explore its integration into LMS platforms, classroom assistants, or mobile applications. At the same time, researchers should pay particular attention to the emotional and motivational experiences of teachers and learners across different quadrant states. Feedback may boost confidence for some learners but create pressure for others. Designing AI prompts that balance encouragement with support will therefore be a crucial priority for the next stage of system development.

As a broader roadmap, this study proposes the notion of a "Bridging Stage": moving from exploratory validation to iterative signal optimization, quadrant-driven prototype testing, and ultimately toward scalable system deployment. This roadmap underscores the continuity between research validation and practical integration.

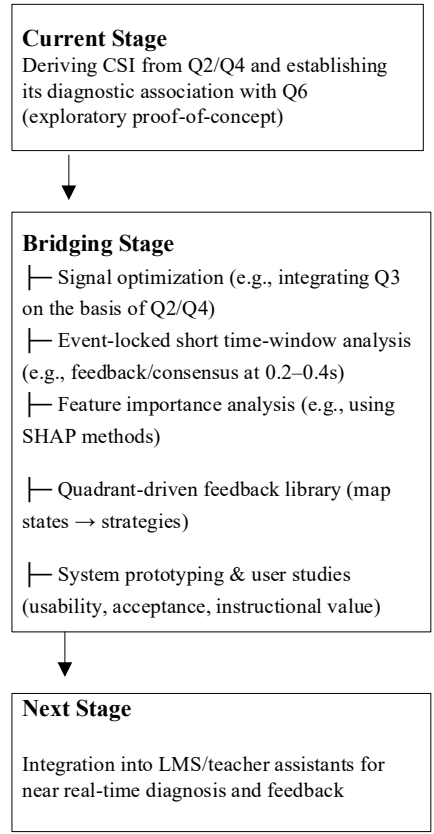

Figure 5.1 Research roadmap

The current stage has centered on constructing the CSI based on Q2 and Q4 and testing its association with sustained motivation (Q6). The next "Bridging Stage" will focus on optimizing signals—such as incorporating Q3 and phase-level micro-signals—while developing a quadrant-based feedback library and conducting small-scale prototypes and user studies. The subsequent stage aims at system-level deployment through integration into LMS platforms or classroom assistants, thereby establishing an AI-ready feedback system.

Summary. The CSI–Q6 quadrant, the prototype design, and the decision-tree model exemplify the key features of AI-readiness: explainable, detectable, and deployable. The study has demonstrated feasibility at the level of classroom tasks while also laying the foundation for future system integration and large-scale application.

Response to research questions. This study also addresses the three research questions posed in Chapter 1. With respect to RQ1, findings show that shared understanding (Q2) and consensus building (Q4) can be successfully operationalized as classroom proxy signals, integrated into the Composite Signal Index (CSI), and shown to be feasible in authentic classroom settings. Regarding RQ2, exploratory analyses revealed a positive association between CSI and sustained motivation (Q6). While larger-scale studies are needed, the initial evidence suggests that the CSI × Q6 quadrant framework can differentiate engagement states and provide diagnostic logic for targeted teacher interventions. Finally, with respect to RQ3, the teacher-facing prototype and quadrant-based prompts demonstrate that these structured signals are not only transparent and interpretable but can also be embedded into a signal–judgment–feedback loop as AI-ready inputs, supporting real-time teacher–AI collaboration and more equitable recognition of learners.

In sum, the exploratory validation provided by this study offers initial answers to the three research questions and, through the positioning of a "Bridging Stage," lays a foundation for future iterations and expansion.

## 6. Conclusion

This study proposed and validated a lightweight, interpretable, and classroom-ready framework for modeling collaborative engagement in technology-enhanced learning environments. By operationalizing neuroscience-informed mechanisms—shared understanding (Q2) and consensus building (Q4)—into observable behavioral proxies and integrating them into a Composite Signal Index (CSI), the study demonstrates how core cognitive–interactive processes can be captured without reliance on intrusive or high-cost multimodal sensing. When combined with sustained motivation (Q6), the CSI supports a quadrant-based diagnostic model that offers transparent and pedagogically meaningful interpretations of group engagement states.

Findings from the authentic ESL classroom show that structured collaborative tasks can reliably elicit observable engagement signals that are theoretically grounded and practically useful. The teacher-facing prototype illustrates one possible way these structured signals may inform **decision support** in technology-enhanced learning settings. Rather than positioning these signals as components of a fully automated system, the design highlights how interpretable indicators can help teachers notice patterns that are difficult to observe in real time. This emphasis on transparency and human judgment is particularly important for recognizing silent or marginalized learners whose engagement may not be reflected in surface behaviors. The study reinforces that silence does not equal disengagement and frequent talk does not guarantee cognitive depth, underscoring the value of mechanism-aligned signals for equitable engagement monitoring.

Beyond its immediate contribution, the framework offers a scalable pathway for future implementation across subjects, grade levels, and learning modalities, including online and hybrid environments. Because Q2, Q4, and Q6 can be captured through low-cost surveys, discourse analysis, or automated transcription, the approach holds promise for integration into learning management systems, lightweight analytics dashboards, and future teacher-informed, AI-supported feedback tools. The proposed "Bridging Stage"—from exploratory validation to iterative refinement and system-level integration—provides a roadmap for advancing engagement modeling as a practical component of technology-enhanced learning.

In summary, this study contributes a transparent, deployable, and equity-oriented framework for understanding collaborative engagement. By bridging learning sciences theory, neuroscience-informed mechanisms, and real-world classroom practice, it lays the groundwork for the next generation of explainable and scalable support systems in collaborative learning.

# References

Acosta, H., Bae, H., Lee, S., Glazewski, K., Hmelo-Silver, C., Mott, B., & Lester, J. (2024). Multimodal learning analytics for predicting student collaboration satisfaction in collaborative game-based learning. In *Proceedings of the 17th International Conference on Educational Data Mining (EDM 2024)* (Paper 19). International Educational Data Mining Society.

https://educationaldatamining.org/edm2024/proceedings/2024.EDM-long-papers.19/index.html

Algumaei, M., Hettiarachchi, I. T., Farghaly, M., & Bhatti, A. (2023). The neuroscience of team dynamics: Exploring neurophysiological measures for assessing team performance. *IEEE Access, 11,* 129173–129194. https://doi.org/10.1109/ACCESS.2023.3332907

Almashour, M., Aldamen, H. A. K., & Jarrah, M. (2025). "They know AI, but they also know us": Student perceptions of EFL teacher identity in AI-enhanced classrooms in Jordan. *Frontiers in Education, 10*, 1611147. https://doi.org/10.3389/feduc.2025.1611147

Apicella, A., Arpaia, P., Frosolone, M., Improta, G., Moccaldi, N., & Pollastro, A. (2022). EEG-based measurement system for monitoring student engagement in learning 4.0. *Scientific Reports, 12*(1), 5857. https://doi.org/10.1038/s41598-022-09578-y

Appleton, J. J., Christenson, S. L., Kim, D., & Reschly, A. L. (2006). Measuring cognitive and psychological engagement: Validation of the Student Engagement Instrument. *Journal of School Psychology, 44*(5), 427–445. https://doi.org/10.1016/j.jsp.2006.04.002

Bach, A., & Thiel, F. (2024). Collaborative online learning in higher education—quality of digital interaction and associations with individual and group-related factors. *Frontiers in Education, 9,* 1156271. https://doi.org/10.3389/feduc.2024.1356271

Biancardi, B., Mancini, M., Ravenet, B., & Varni, G. (2024). Modelling the "transactive memory system" in multimodal multiparty interactions. *Journal on Multimodal User Interfaces, 18*(1), 103–117. https://doi.org/10.1007/s12193-023-00426-5

Bosch, N., D'Mello, S. K., Ocumpaugh, J., Baker, R. S., & Shute, V. (2016). Using video to automatically detect learner affect in computer-enabled classrooms. *ACM Transactions on Interactive Intelligent Systems (TiiS), 6*(2), Article 17, 1–26. https://doi.org/10.1145/2946837

Castro, M. S. O., Mello, R. F., Viterbo, O., Fiorentino, G., Spikol, D., Bano, M., & Gašević, D. (2023). Understanding peer feedback contributions using natural language processing. In O. Viberg, A. Mozelius,

M. Scheffel, & M. Li (Eds.), *Proceedings of the 18th European Conference on Technology Enhanced Learning (EC-TEL 2023)* (Lecture Notes in Computer Science, Vol. 14200, pp. 399–414). Springer. https://doi.org/10.1007/978-3-031-42682-7_27

Chinta, S. V., Wang, Z., Yin, Z., Hoang, N., Gonzalez, M., Le Quy, T., & Zhang, W. (2024). FairAIED: Navigating fairness, bias, and ethics in educational AI applications. *arXiv preprint arXiv:2407.18745*. https://doi.org/10.48550/arXiv.2407.18745

Davidesco, I., Laurent, E., Valk, H., West, T., Milne, C., Poeppel, D., & Dikker, S. (2023). The temporal dynamics of brain-to-brain synchrony between students and teachers predict learning outcomes. *Psychological Science, 34*(5), 633–643. https://doi.org/10.1177/09567976231163872

D'Mello, S. K., Dieterle, E., & Duckworth, A. (2017). Advanced, analytic, automated (AAA) measurement of engagement during learning. *Educational Psychologist, 52*(2), 104–123. https://doi.org/10.1080/00461520.2017.1281747

Dikker, S., Wan, L., Davidesco, I., Kaggen, L., Oostrik, M., McClintock, J., Rowland, J., Michalareas, G., Van Bavel, J. J., Ding, M., & Poeppel, D. (2017). Brain-to-brain synchrony tracks real-world dynamic group interactions in the classroom. *Current Biology, 27*(9), 1375–1380. https://doi.org/10.1016/j.cub.2017.04.002

Ding, D., & Yusof, A. M. B. (2025). Investigating the role of AI-powered conversation bots in enhancing L2 speaking skills and reducing speaking anxiety: A mixed methods study. *Humanities and Social Sciences Communications, 12*(1), Article 155. https://doi.org/10.1057/s41599-025-05550-z

Du, X., Zhang, L., Hung, J.-L., Li, H., Tang, H., & Xie, Y. (2022). Understand group interaction and cognitive state in online collaborative problem solving: Leveraging brain-to-brain synchrony data. *International Journal of Educational Technology in Higher Education, 19*, 52. https://doi.org/10.1186/s41239-022-00356-4

Fredricks, J. A., Blumenfeld, P. C., & Paris, A. H. (2004). School engagement: Potential of the concept, state of the evidence. *Review of Educational Research, 74*(1), 59–109. https://doi.org/10.3102/00346543074001059

Gao, R., Chen, M.-B., Frermann, L., & Lau, J. H. (2025, February 24). *Moderation matters: Measuring conversational moderation impact in English as a second language group discussion* [Preprint]. *arXiv*. https://doi.org/10.48550/arXiv.2502.18341

Goodwin, L. (2022). *Educational neuroscience, artificial intelligence, and technology in higher education* (Publication No. 29254100) [Doctoral dissertation, University of Pennsylvania]. *ProQuest Dissertations Publishing.*

Gu, C., Peng, Y., Nastase, S. A., Mayer, R. E., & Li, P. (2024). Onscreen presence of instructors in video lectures affects learners' neural synchrony and visual attention during multimedia learning. *Proceedings of the National Academy of Sciences, 121*(12), e2309054121. https://doi.org/10.1073/pnas.2309054121

Guerrero-Sosa, J. D. T., Romero, F. P., Menéndez-Domínguez, V. H., Serrano-Guerrero, J., Montoro-Montarroso, A., & Olivas, J. A. (2025). A comprehensive review of multimodal analysis in education. *Applied Sciences, 15*(11), 5896. https://doi.org/10.3390/app15115896

Gupta, S., Kumar, P., & Tekchandani, R. (2024). Artificial intelligence-based cognitive state prediction in an e-learning environment using multimodal data. *Multimedia Tools and Applications, 83*(1), 64467–64498. https://doi.org/10.1007/s11042-023-18021-x

Hadwin, A. F., Järvelä, S., & Miller, M. (2018). Self-regulation, co-regulation, and shared regulation in collaborative learning environments. In D. Schunk & J. Greene (Eds.), *Handbook of self-regulation of learning and performance* (2nd ed., pp. 83–106). Routledge.

Hasson, U., Ghazanfar, A. A., Galantucci, B., Garrod, S., & Keysers, C. (2012). Brain-to-brain coupling: A mechanism for creating and sharing a social world. *Trends in Cognitive Sciences, 16*(2), 114–121. https://doi.org/10.1016/j.tics.2011.12.007

Henrie, C. R., Halverson, L. R., & Graham, C. R. (2015). Measuring student engagement in technology-mediated learning: A review. *Computers & Education, 90,* 36–53. https://doi.org/10.1016/j.compedu.2015.09.005

Hildebrandt, T., & Mehnen, L. (2024). Cross-course process mining of student clickstream data: Aggregation and group comparison. *Proceedings of the 2024 26th International Conference on Business Informatics (CBI)*. IEEE. https://doi.org/10.1109/CBI62504.2024.00020

Holstein, K., & Doroudi, S. (2021). Equity and artificial intelligence in education: Will "AIEd" amplify or alleviate inequities in education? *arXiv preprint arXiv:2104.12920.* https://doi.org/10.48550/arXiv.2104.12920

Jamil, N., & Belkacem, A. N. (2024). Advancing real-time remote learning: A novel paradigm for cognitive enhancement using EEG and eye-tracking analytics. *IEEE Access, 12,* 3422926. https://doi.org/10.1109/ACCESS.2024.3422926

Järvelä, S., Nguyen, A., Vuorenmaa, E., Malmberg, J., & Järvenoja, H. (2023). Predicting regulatory activities for socially shared regulation to optimize collaborative learning. *Computers in Human Behavior, 144*, 107737. https://doi.org/10.1016/j.chb.2023.107737

Jeitziner, L. T., Paneth, L., Rack, O., Bleisch, S., & Zahn, C. (2025). Measuring the quality of collaborative group engagement: Development and validation of the QCGE self-assessment scale (QCGE-SAS). *International Journal of Computer-Supported Collaborative Learning, 20*(1), 101–133. https://doi.org/10.1007/s11412-025-09445-8

Johnston, L. J., Griffin, J. E., Manolopoulou, I., & Jendoubi, T. (2024). Uncovering student engagement patterns in Moodle with interpretable machine learning. *arXiv preprint arXiv:2412.11826.* https://doi.org/10.48550/arXiv.2412.11826

Karimah, S. N., & Hasegawa, S. (2022). Automatic engagement estimation in smart education/learning settings: A systematic review of engagement definitions, datasets, and methods. *International Journal of Educational Technology in Higher Education, 19*(1), 1–25. https://doi.org/10.1186/s40561-022-00212-y

Kovari, A. (2025). A systematic review of AI-powered collaborative learning in higher education: Trends and outcomes from the last decade. *Social Sciences & Humanities Open, 11*, 101335. https://doi.org/10.1016/j.ssaho.2025.101335

Kuhlman, S. L., Plumley, R., Evans, Z., Bernacki, M. L., Greene, J. A., Hogan, K. A., Berro, M., Gates, K., & Panter, A. (2024). Students' active cognitive engagement with instructional videos predicts STEM learning. *Computers & Education, 216*, 105050. https://doi.org/10.1016/j.compedu.2024.105050

Lawrence, L. A. Aoyama, & Weinberger, A. (2022). Being in-sync: A multimodal framework on the emotional and cognitive synchronization of collaborative learners. *Frontiers in Education, 7*. https://doi.org/10.3389/feduc.2022.867186

Markiewicz, R., Segaert, K., & Mazaheri, A. (2024). Brain-to-brain coupling forecasts future joint action outcomes. *iScience, 27*(2), 110802. https://doi.org/10.1016/j.isci.2024.110802

Michinov, N., & Michinov, E. (2009). Investigating the relationship between transactive memory and performance in collaborative learning. *Learning and Instruction, 19*(1), 43–54. https://doi.org/10.1016/j.learninstruc.2008.01.003

Nasir, J., Kothiyal, A., Bruno, B., & Dillenbourg, P. (2022). Many are the ways to learn: Identifying multimodal behavioral profiles of collaborative learning in constructivist activities. *International Journal*

*of Computer-Supported Collaborative Learning, 16*(4), 485–523. https://doi.org/10.1007/s11412-021-09358-2

Nennig, H. T., Cole, R., States, N. E., Macrie-Shuck, M., Fateh, S., & Gunes, Z. D. K. (2023). Exploring social and cognitive engagement in small groups through a community of learners (CoL) lens. *Chemistry Education Research and Practice, 24*(6), 1077–1099. https://doi.org/10.1039/D3RP00071K

Pan, Y., Cheng, X., & Hu, Y. (2023). Three heads are better than one: Cooperative learning brains wire together when a consensus is reached. *Cerebral Cortex, 33*(5), 1155–1169. https://doi.org/10.1093/cercor/bhac127

Pan, Y., Dikker, S., Goldstein, P., Zhu, Y., Yang, C., & Hu, Y. (2020). Instructor–learner brain coupling discriminates between instructional approaches and predicts learning. *NeuroImage, 211*, 116657. https://doi.org/10.1016/j.neuroimage.2020.116657

Pan, Y., Novembre, G., Song, B., Li, X., & Hu, Y. (2018). Interpersonal synchronization of inferior frontal cortices tracks social interactive learning of a song. *NeuroImage, 183*, 280–290. https://doi.org/10.1016/j.neuroimage.2018.08.005

Pan, Y., Novembre, G., Song, B., Zhu, Y., & Hu, Y. (2021). Dual brain stimulation enhances interpersonal learning through spontaneous movement synchrony. *Social Cognitive and Affective Neuroscience, 16*(1–2), 210–221. https://doi.org/10.1093/scan/nsaa080

Panadero, E., Jonsson, A., Pinedo, L., & Fernández-Castilla, B. (2023). Effects of rubrics on academic performance, self-regulated learning, and self-efficacy: A meta-analytic review. *Educational Psychology Review, 35*, 113. https://doi.org/10.1007/s10648-023-09823-4

Park, S., Nixon, N., D'Mello, S., Shariff, D., & Choi, J. (2025). Understanding collaborative learning processes and outcomes through student discourse dynamics. *In LAK '25: Proceedings of the 15th International Learning Analytics and Knowledge Conference* (pp. 938–943). Association for Computing Machinery. https://doi.org/10.1145/3706468.3706547

Peltoniemi, A. J., Lämsä, J., Lehesvuori, S., & Hämäläinen, R. (2025). Understanding the role of I-positions facilitating knowledge construction in a computer-supported collaborative learning environment. *International Journal of Computer-Supported Collaborative Learning, 20*(2), 245–269. https://doi.org/10.1007/s11412-025-09447-6

Reinero, D. A., Dikker, S., & Van Bavel, J. J. (2021). Inter-brain synchrony in teams predicts collective performance. *Social Cognitive and Affective Neuroscience, 16*(1–2), 43–57. https://doi.org/10.1093/scan/nsaa135

Schilbach, L., & Redcay, E. (2025). Synchrony across brains. *Annual Review of Psychology, 76*, 883–911. https://doi.org/10.1146/annurev-psych-080123-101149

Sharma, P., Stewart, A. E. B., Li, Q., Ravichander, K., & Walker, E. (2024). Building learner activity models from log data using sequence mapping and hidden Markov models. In *Proceedings of the 17th International Conference on Educational Data Mining (EDM 2024)* (Paper 60). International Educational Data Mining Society. https://educationaldatamining.org/edm2024/proceedings/2024.EDM-short-papers.60/index.html

Sievers, B., Welker, C., Hasson, U., Kleinbaum, A. M., & Wheatley, T. (2024). Consensus-building conversation leads to neural alignment. *Nature Communications, 15,* 43253. https://doi.org/10.1038/s41467-023-43253-8

Stephens, G. J., Silbert, L. J., & Hasson, U. (2010). Speaker–listener neural coupling underlies successful communication. *Proceedings of the National Academy of Sciences, 107*(32), 14425–14430. https://doi.org/10.1073/pnas.1008662107

Thurn, C. M., Edelsbrunner, P. A., Berkowitz, M., Deiglmayr, A., & Schalk, L. (2023). Questioning central assumptions of the ICAP framework. *npj Science of Learning, 8*(1), 49. https://doi.org/10.1038/s41539-023-00197-4

Volet, S., Vauras, M., & Salonen, P. (2009). Self- and social regulation in learning contexts: An integrative perspective. *Educational Psychologist, 44*(4), 215–226. https://doi.org/10.1080/00461520903213584

Xing, W., Zhu, G., Arslan, O., Shim, J., & Popov, V. (2022). Using learning analytics to explore the multifaceted engagement in collaborative learning. *Journal of Computing in Higher Education, 35,* 633–662. https://doi.org/10.1007/s12528-022-09343-0

Xu, B., Stephens, J. M. S., & Lee, K. (2024). Assessing student engagement in collaborative learning: Development and validation of new measure in China. *Asia-Pacific Education Researcher, 33*(3), 395–405. https://doi.org/10.1007/s40299-023-00737-x

Zheng, L., Long, M., Niu, J., & Zhong, L. (2023). An automated group learning engagement analysis and feedback approach to promoting collaborative knowledge building, group performance, and socially shared regulation in CSCL. *International Journal of Computer-Supported Collaborative Learning, 18*(1), 101–133. https://doi.org/10.1007/s11412-023-09386-0

# Declarations

### Ethics statement

This work was conducted in an established ESL class with adult learners. The activity involved only normal educational practices (a structured reading–retelling–consensus task embedded in regular instruction); no intervention beyond routine teaching was introduced. Analyses used fully de-identified questionnaire ratings and classroom notes; no names, faces, audio/video identifiers, or contact information were collected or stored. Under the U.S. Common Rule 45 CFR 46.104(d)(1) (normal educational practices), IRB review was not required. All procedures conformed to standard publishing ethics.

### Consent to participate / publication

Not applicable; the study used de-identified classroom data and includes no identifiable images or personal data.

### Funding

This research did not receive any specific grant from funding agencies in the public, commercial, or not-for-profit sectors.

### Declaration of interest

The authors declare that they have no known competing financial interests or personal relationships that could have appeared to influence the work reported in this paper.

### Declaration of generative AI and AI-assisted technologies in the writing process

During the preparation of this work, the author(s) used ChatGPT (OpenAI) to improve readability and language clarity. No AI tools were used for analysis, data processing, or interpretation. After using this tool, the author(s) reviewed and edited the content as needed and take(s) full responsibility for the content of the publication.

# Appendix A. Task Materials and Survey Items

**A.1 Collaborative Speaking Task Sheet**

# Collaborative Speaking Task Sheet

Goal: Practice speaking, listening, and peer collaboration using two short stories.

Group Size: 2–3 students     Total Time: About 20–25 minutes (10–12 minutes per story)

Round 1 – Story 1: Darwin's Theory

1. **Turn-taking Reading**

   – Read one sentence each.

   – Help each other with pronunciation or meaning.

2. **Individual Retelling**

   – Each student retells the story in their own words.

   – Focus on the main ideas — it doesn't have to be perfect.

   – You can give short feedback to help your partner improve:

• "You missed…" – a part they forgot    • "Maybe you can say…" – a better way to say it

• "Try adding…" – extra detail to make it clearer

3. **Group Consensus Discussion**

   – Discuss: Do you think Darwin's theory still matters today? Why or why not?

   – Try to agree as a group, or explain your different opinions.

   –  Write one sentence as a group answer.

Round 2 – Story 2: Natural Selection

1. Turn-taking Reading

   – Read one sentence each.

   – Help each other with new words or ideas.

2. Individual Retelling

   – Each student retells the story in their own words.

   – Focus on key points and clarity.

3. Group Consensus Discussion

   – Discuss: Do you think natural selection still happens today? Can you give an example?

   – Share your ideas and try to reach agreement.

   –  Write one sentence as a group answer.

**A.2 Student Self-Evaluation Form (Survey Items)**

# Student Self-Evaluation Form

(To be completed after today's speaking task)

Instructions: Please reflect on your experience in today's group activity.

**<u>After Story 1</u> (**Please answer this question right after Story 1 — before moving to Story 2.**)**

Check the box that best shows how much you agree with each sentence.

1 = Strongly Disagree    5 = Strongly Agree

| No. | Statement | 1 | 2 | 3 | 4 | 5 |
|-----|-----------|---|---|---|---|---|
| 1 | Today's activity made me more willing to speak English. | ☐ | ☐ | ☐ | ☐ | ☐ |

**<u>After Story 2</u>** (You may now complete the rest of the form after Story 2.)

Check the box that best shows how much you agree:

1 = Strongly Disagree    5 = Strongly Agree

| No. | Statement | 1 | 2 | 3 | 4 | 5 |
|-----|-----------|---|---|---|---|---|
| 1 | Today's activity made me more willing to speak English. | ☐ | ☐ | ☐ | ☐ | ☐ |
| 2 | I understood most of what my group members said. | ☐ | ☐ | ☐ | ☐ | ☐ |
| 3 | Giving and receiving feedback helped me improve my ideas. | ☐ | ☐ | ☐ | ☐ | ☐ |
| 4 | Group discussion helped us reach agreement. | ☐ | ☐ | ☐ | ☐ | ☐ |
| 5 | I felt more confident sharing my opinion in a group. | ☐ | ☐ | ☐ | ☐ | ☐ |
| 6 | I would like to do this kind of activity again. | ☐ | | ☐ | ☐ | ☐ | ☐ |

**Optional: What part of today's activity helped you the most?**

_______________________________________________

_______________________________________________

_______________________________________________